\documentclass[twocolumn,showpacs,a4,prb]{revtex4}
\usepackage{graphicx}%
\usepackage{dcolumn}
\usepackage{bm}
\makeatletter
\def\btt#1{\texttt{\@backslashchar#1}}%
\DeclareRobustCommand\bblash{\btt{\@backslashchar}}%
\makeatother

\begin{document}

\preprint{Carbon-doped.TEX}

\title{$^{11}$B NMR study of pure and lightly carbon doped MgB$_2$ superconductors}

\author{M. Karayanni, G. Papavassiliou, M. Pissas, and M. Fardis}
\affiliation{Institute of Materials Science, NCSR,  Demokritos, 153 10 Aghia Paraskevi, Athens, Greece}
\author{K. Papagelis and K. Prassides}
\affiliation{Department of Chemistry, University of Sussex, Brighton BN1 9QJ, UK}
\author{T. Takenobu and Y. Iwasa}
\affiliation{Institute for Materials Research, Tohoku University - Sendai 980-8577, Japan and
CREST, Japan Science and Technology Corporation - Kawaguchi 332-0012, Japan} 
\date{\today }

\begin{abstract}
We report a $^{11}$B NMR line shape and spin-lattice relaxation rate ($1/(T_1T)$) study of pure and lightly carbon doped MgB$_{2-x}$C$_{x}$ for $x=0$, $0.02$, and $0.04$, in the vortex state and in magnetic field of $23.5$ kOe. We show that while pure MgB$_2$ exhibits the magnetic field distribution from superposition of the normal and the Abrikosov state, slight replacement of boron with carbon unveils the magnetic field distribution of the pure Abrikosov state. This indicates a considerable increase of $H_{c2}^c$ with carbon doping with respect to pure MgB$_2$. The spin-lattice relaxation rate $1/(T_1T)$ demonstrates clearly the presence of a coherence peak right below $T_c$ in pure MgB$_2$, followed by a typical BCS decrease on cooling. However, at temperatures lower than $\approx 10$K strong deviation from the BCS behavior is observed, probably from residual contribution of the vortex dynamics. In the carbon doped systems both the coherence peak and the BCS temperature dependence of $1/(T_1T)$ weaken, an effect attributed to the gradual shrinking of the $\sigma $ hole cylinders of the Fermi surface with electron doping.  
\end{abstract}
\maketitle

\section{Introduction}
The recent discovery of superconductivity in MgB$_2$ \cite{Nagamatsu01} has motivated extensive experimental and theoretical work, as this material has the highest critical temperature among superconducting binary compounds (T$_c\approx 40$K). MgB$_2$ is isostructural and isoelectronic with intercalated graphite, and has a simple AlB$_2$-type crystal structure (space group $P6/mmm$) consisting of alternating close-packed Mg layers and graphite-like honeycomb B layers. Most theoretical and experimental work on MgB$_2$ has been devoted to the understanding of the mechanism responsible for its unusual superconductivity. By now it is widely accepted that MgB$_2$ is a non-conventional BCS-type superconductor. First-principle calculations \cite{Liu01,Choi02,Koshelev03} show that superconductivity resides in two groups of bands: (i) the two strongly superconducting $\sigma $ bands and (ii) the two weakly superconducting $\pi $ bands. A large number of experiments, in particular studies with Raman spectroscopy \cite{Chen01}, scanning tunneling microscopy \cite{Giubileo01,Bollinger01,Iavarone02,Eskildsen02}, muon spin resonance 
\cite{Niedermayer02,Papagelis03a,Papagelis03b}, point-contact spectroscopy \cite{Szabo01,Gonnelli03,Yanson04,Bugoslavsky04} and specific-heat \cite{Bouquet01} measurements support the notion that MgB$_2$ is a double gap superconductor. On the basis of this evidence new physics has been unveiled. For example, theoretical \cite{Kogan02,Miranovic03} and experimental \cite{Cubitt03,Lyard03} works have shown that the anisotropy of the upper critical field $\gamma_\xi =H_{c2}^{ab}/H_{c2}^c\approx 6$ differs substantially from the anisotropy of the London penetration depth $\gamma_\lambda  =\lambda ^c/\lambda ^{ab}\approx 1.2$, at low temperatures. It has been also suggested that MgB$_2$ posseses  a peculiar vortex structure \cite{Koshelev03,Eskildsen02}, where vortices in the $\pi $ band are characterized by absence of localized states in the core, very large vortex core size, and strong core overlap.

Many of the efforts have been focused in the study of doped MgB$_2$ by partially substituting Mg or B sites with electron or hole dopants. Such substituents modify (i) the crystal structure, and (ii) the electron density of states at the Fermi level $N(E_F)$, and therefore are extremely important in analyzing the band structure and the microscopic nature of superconductivity in MgB$_2$. A systematic study of Al$_x$Mg$_{1-x}$B$_2$ samples with Mg replaced by Al has already been undertaken by several groups. \cite{Slusky01,Pissas02,Papavassiliou02,Serventi03} In particular, $^{27}$Al and $^{11}$B NMR measurements in Al$_x$Mg$_{1-x}$B$_2$ \cite{Papavassiliou02,Kotegawa02,Serventi03}, provided strong evidence that the Fermi surface is made of hole-type $\sigma $-bonding $2D$ cylindrical sheets, which fill up upon Al doping and collapse at $x\approx 0.55$, together with hole-type and electron-type $3D$ $\pi $-bonding tubular networks, in accordance with theoretical predictions. \cite{Belashchenko01} These results suggest anisotropic pairing and multigap superconductivity for MgB$_2$. 
An alternative approach is to introduce dopants into the B sublattice. Several groups have attempted to replace B with carbon. \cite{Papagelis03a,Papagelis03b,Takenobu01,Lee03,Avdeev03,Ribeiro03,Mickelson02,Bharathi02,Cheng02,Paranthaman01,Maurin02,Maurin02b,Papagelis02, Arvanitidis} The results reported in the literature indicate that the degree of carbon doping depends on the procedure and the starting material used in the synthesis of MgB$_{2-x}$C$_x$. Similarily with Al$_x$Mg$_{1-x}$B$_2$, the replacement of boron with carbon leads again to doping with one electron, and a significant depression of $T_c$ \cite{Takenobu01,Lee03,Avdeev03, Mickelson02}, although the effect is more pronounced in the case of MgB$_{2-x}$C$_x$. At the same time, x-ray measurements have shown that MgB$_{2-x}$C$_x$ exhibits a highly anisotropic lattice contraction with the lattice parameter $a$ decreasing almost linearly with increasing doping level \cite{Takenobu01,Lee03,Avdeev03,Mickelson02, Maurin02b}, while the $c$ parameter is virtually unaffected by the doping level, indicating that carbon is exclusively substituted in the boron honeycomb layer without affecting the interlayer interactions. Besides, data taken from tunneling \cite{Schmidt03}, specific heat \cite{Ribeiro03},  point-contact \cite{Samuely03}, and muon SR measurements \cite{Papagelis03a,Papagelis03b} reveal that despite the homogeneous and random nature of C substitution the two-gap behavior is preserved in MgB$_{2-x}$C$_x$. This is in contrast to theoretical predictions, according to which enhanced quasiparticle interband scattering should lead to an effective one-gap behavior in the doped samples \cite{Liu01} and suggests that the interband scattering is selectively suppressed in MgB$_{2-x}$C$_x$. \cite{Schmidt03}

In order to elucidate further the complicated electronic structure of MgB$_2$, we have performed a comparative $^{11}$B NMR study on pure MgB$_2$ and  carbon doped MgB$_{2-x}$C$_x$ samples for $x=0.0$, $0.02$, and $0.04$. Our results indicate a considerable increase of $H_{c2}^c$ and strong supression of both the coherence peak and the BCS temperature dependence of $1/(T_1T)$ with carbon doping. These results are in agreement with recent ac-susceptibility and torque measurements \cite{Pissas03,Ohmichi03}, which show strong enhancement of $H_{c2}^c$, passage from the clean to the dirty limit regime, and shrinking of the $\sigma $ band hole-type parts of the Fermi surface with carbon doping.

\section{Sample preparation and characterization}

Polycrystalline pure MgB$_2$ and carbon doped MgB$_{2-x}$C$_x$ samples for $x=0.02$ and $0.04$ were prepared by reaction of Mg, amorphous B and carbon black at $900$K for $2$ h, as described in ref. \cite{Takenobu01,Papagelis02} Synchrotron X-ray diffraction patterns, Raman and ac-susceptibility measurements of samples from the same batch are presented elsewere. \cite{Takenobu01,Papagelis02, Maurin02, Maurin02b, Arvanitidis} $^{11}$B NMR line shape measurements of the central transition ($-1/2\rightarrow 1/2$) were performed on a home-built spectrometer operating in external magnetic field $H_{\rm o}=23.5$ kOe. The spectra were obtained from the Fourier transform of half of the echo, following a typical $\pi/2-\tau-\pi$ spin-echo pulse sequence. \cite{Papavassiliou01} The irradiation frequency was the same in all experiments. The spectral bandwidth around the irradiation frequency was $\approx 150$ kHz, which is enough to cover adequately the NMR signals at low temperatures. The $^{11}$B $T_1$ of the central line was determined by applying a saturation recovery technique, and fitting with the two exponential relaxation function that is appropriate for $I=3/2$ nuclei, $m(t)=(M(\infty )-M(t))/M(\infty) =0.1exp(-t/T_1)+0.9exp(-6t/T_1)$. \cite{Andrew61}

\section{$^{11}$B NMR measurements}

\subsection{$^{11}$B NMR line shapes and the mixed state}

When a type - II superconductor is placed into an external magnetic field $H_{c1}<H_{\rm o}<H_{c2}$, at temperatures lower than the temperature of the second critical field $T_{c2}$, the magnetic field penetrates partially in the form of thin filaments of flux, and generates a vortex lattice with a characteristic magnetic field distribution $f(B)$. This can be excellently monitored by using magnetic resonance techniques, like muon spectroscopy (for a review see ref. \cite{Sonier00}) and NMR \cite{Fite66,Lee00,Papavassiliou01}, as the resonance frequency is directly proportional to the local magnetic field. In case of an ideal hexagonal vortex lattice, a characteristic frequency distribution is formed with a peak corresponding to the saddle point of the magnetic field distribution, which is located midway between two vortices, and two steps at the maximum ($H_{{\rm max}}$) and minimum ($H_{{\rm min}}$) fields \cite{Fite66,Brandt}, corresponding to the vortex core and the center of a vortex triangle, respectively. This distribution depends on the two fundamental length scales of superconductivity: the magnetic penetration length $\lambda$ (the variance of the frequency distribution is inversely proportional to $\lambda$), and the coherence length $\xi$, which defines the "extension" of a vortex. \cite{Sonier00,Brandt,Niedermayer99} It is thus possible from the NMR line shapes to acquire information about the topology and dynamics of the vortex lattice, as well as parameters that reflect the symmetry of the superconducting pairing.

\begin{figure}[htbp] \centering
\includegraphics[angle=0,width=8.5cm]{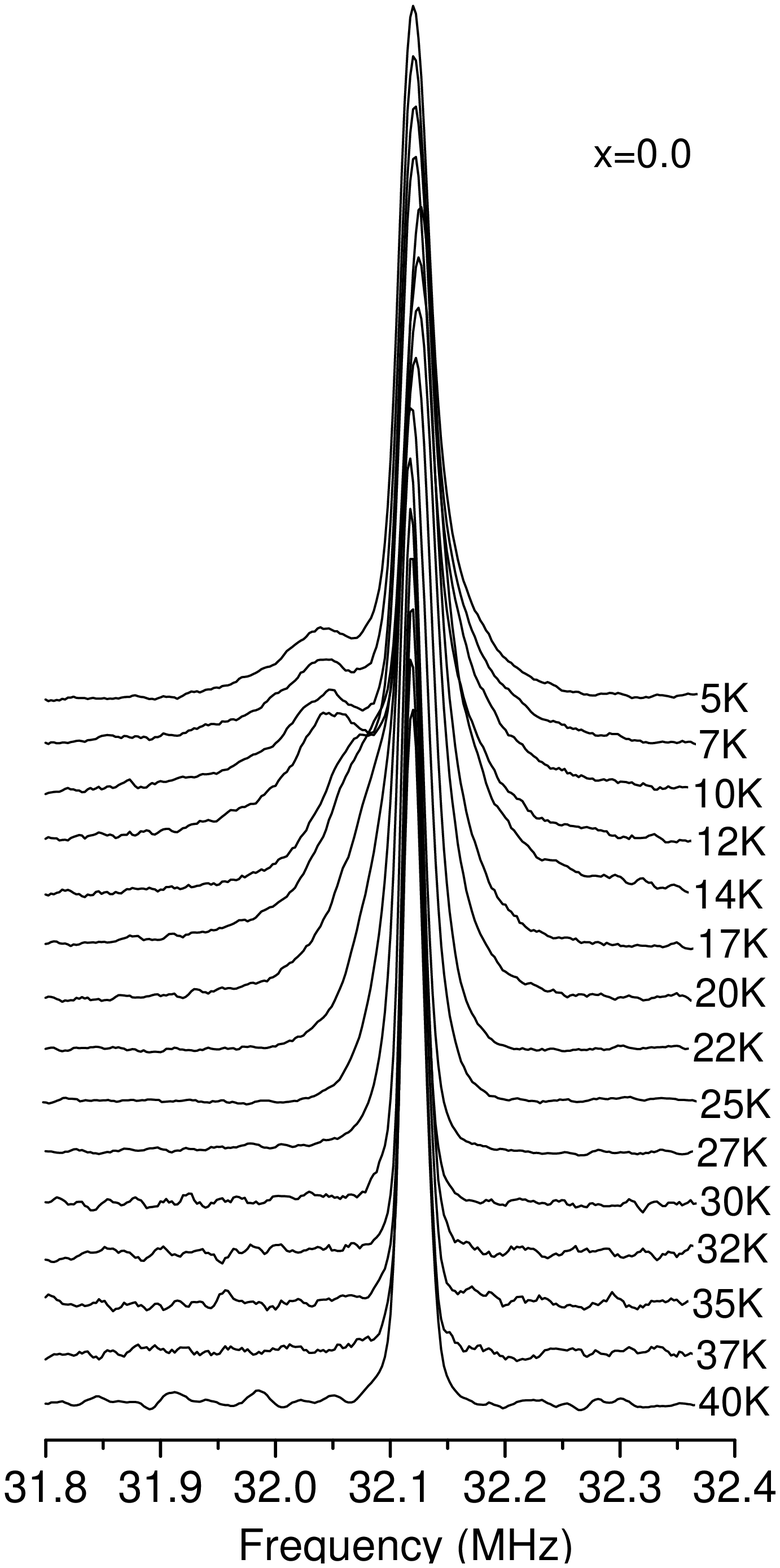}
\caption{$^{11}$B NMR line shapes as a function of temperature for pure MgB$_2$ under a magnetic field $H=23.5$ kOe. For clarity, each spectrum is normalized to its maximum intensity.}
\label{fig1}%
\end{figure}

\begin{figure}[tbp] \centering
\includegraphics[angle=0,width=8.5cm]{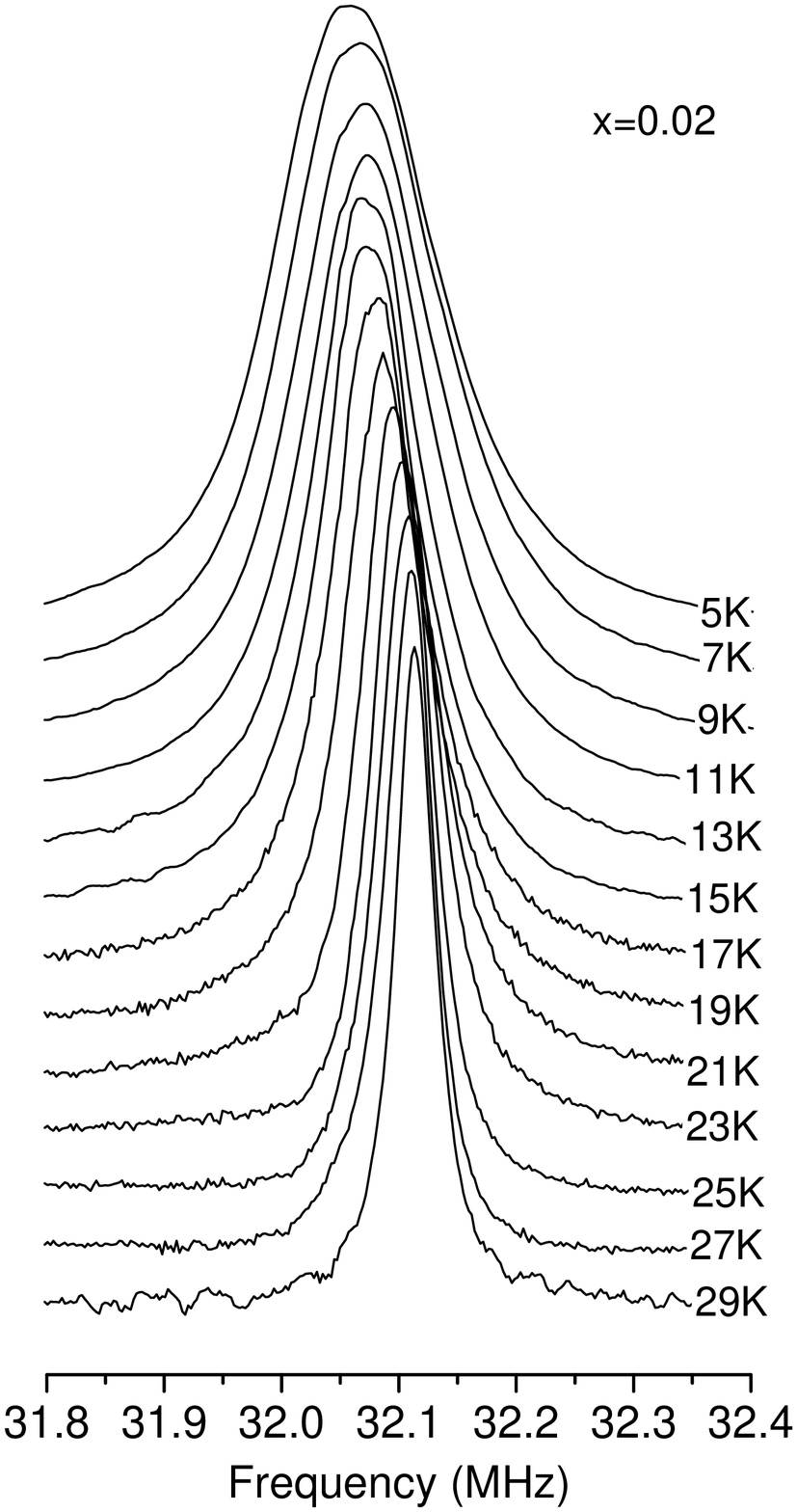}
\caption{
$^{11}$B NMR line shapes of MgB$_{1.98}$C$_{0.02}$ as a function of temperature, under a magnetic field $H=23.5$ kOe. For clarity, each spectrum is normalized to its maximum intensity.}
\label{fig2}
\end{figure}

\begin{figure}[tbp] \centering
\includegraphics[angle=0,width=8.5cm]{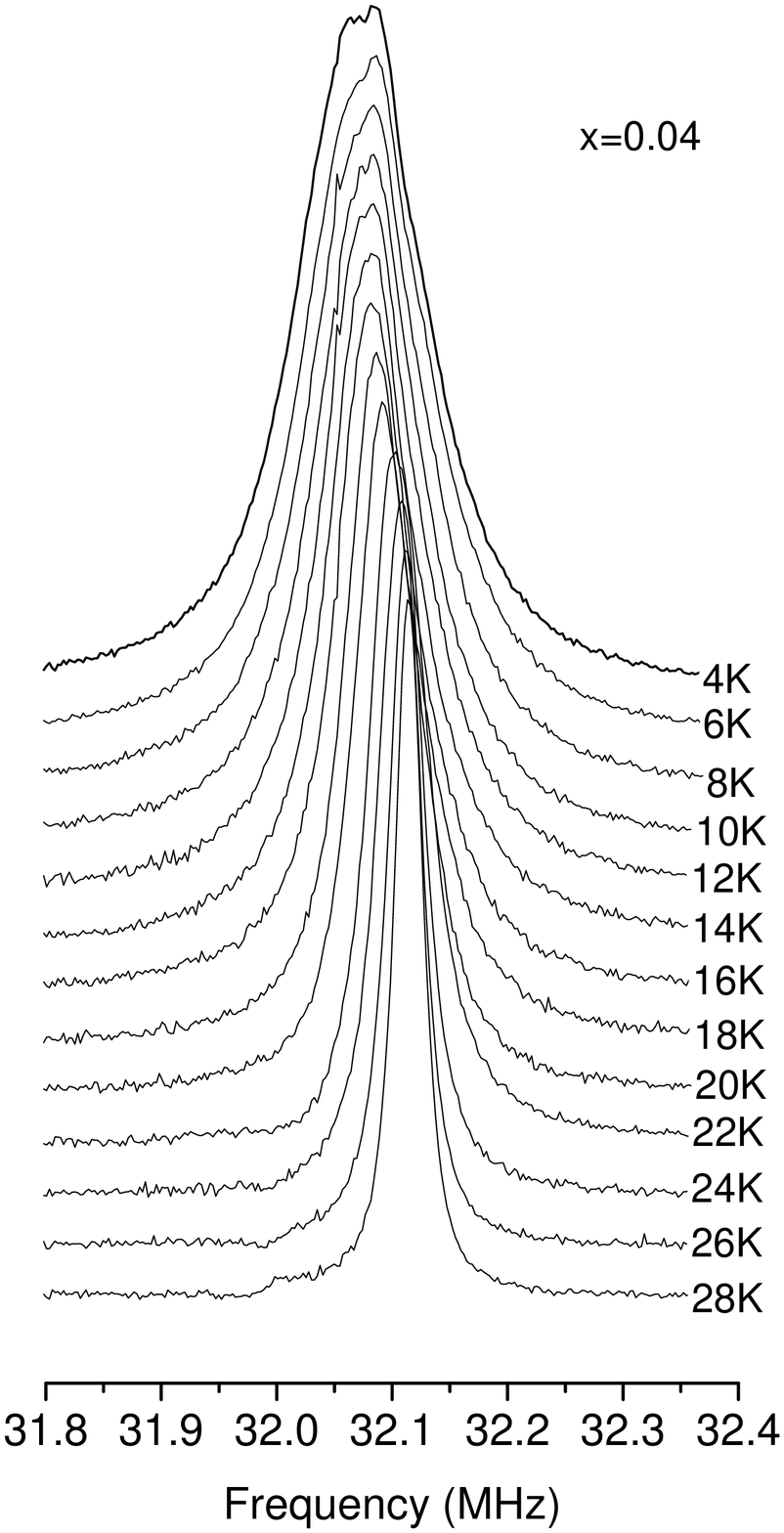}
\caption{
$^{11}$B NMR line shapes of MgB$_{1.96}$C$_{0.04}$ as a function of temperature, under a magnetic field $H=23.5$ kOe. For clarity, each spectrum is normalized to its maximum intensity.}
\label{fig3}
\end{figure}

The $^{11}$B NMR spectra of the investigated pure and carbon doped MgB$_2$ powder samples, in a magnetic field of $23.5$ kOe, are shown as a function of temperature in Figures \ref{fig1}, \ref{fig2}, and \ref{fig3}. In the case of pure MgB$_2$, the line shapes in the normal state remain unchanged down to $T_{c2}$, while for $T<T_{c2}$ a new peak develops at the low frequency side. In a recent work, the direct comparison of the $^{11}$B NMR line shapes with dc-magnetic susceptibility measurements has shown that the frequency position and the shape of the low frequency peak by decreasing temperature, follows the development of the vortex lattice. \cite{Papavassiliou01} In view of this fact, the persistence of the normal state signal down to the lowest measured temperature $T=5$K, has been explained as manifestation of the high anisotropy in the upper critical field $H_{c2}$. \cite{Papavassiliou01,Simon01} Considering that in pure MgB$_2$ the anisotropy ratio is equal to $\gamma\approx 6$ at low temperatures \cite{Simon01,Angst02,Cubitt03}, it is expected that in external magnetic field of $23.5$ kOe ($H_{c2}^c < 2.35$ Tesla) a part of the randomly oriented grains remains in the normal state (unshifted peak), while the rest is in the vortex state. This gives rise to superposition of NMR signals from the vortex and the normal states. The above argument becomes obvious if we consider that the upper critical field in each crystallite varies with the angle between the external magnetic field and the $c-$axis as, $H_{{\rm c2}}(\theta)=H_{\rm c2}^{\rm ab}(1+(\gamma_{\xi }^2-1) \cos^2\theta)^{-1/2}$. According to this formula, only crystallites with $H_{\rm o}<H_{\rm c2}(\theta)$ would give the characteristic signal of a type-II superconductor in the vortex state. If for a part of the grains, $H_{\rm o}>H_{\rm c2}(\theta)$, then the NMR line shape would be the sum of spectra coming from crystallites in the normal and the vortex state. Consequently, for $H_{\rm o}<H_{\rm c2}^c$, the NMR line shapes will reflect solely the magnetic field distribution of the pure vortex state, whereas for $H_{\rm c2}^{ab}<H_0$ only the line shape of the normal state will be observed. At this point we must note that the polycrystaline samples show a lower $H_{c2}^c$, in comparison to single crystals, where $H_{c2}^c\geq 30$ kOe has been observed. This fact may be explained as showing that polycrystalline samples are more clean than single crystals and therefore exhibit a higher $\xi _{ab}$. \cite{Pissas03,Ohmichi03}

It is thus quite remarkable that in the carbon-doped systems the normal state signal component disappears below $T_{c2}$, as clearly observed in Figures \ref{fig2} and \ref{fig3}. In particular, in the case of MgB$_{1.98}$C$_{0.02}$, the line shape starts to broaden below $\approx 26-27$K, while shifting to lower frequencies. The disappearance of the normal state signal component indicates an enhancement (in comparison with pure MgB$_2$) of $H^c_{c2}$ above $23.5$ kOe, thus giving rise to the pure vortex state at low temperatures. However, in the case of MgB$_{1.96}$C$_{0.04}$, line shapes in the mixed state appear to be quite more symmetric, and less shifted than the line shapes of the $x=0.02$ sample, which is a signature of the gradual deformation of the vortex lattice by increasing carbon doping. Similar increase of $H_{c2}^c$ and reduction of the anisotropy have been previously observed in the mixed superconducting state of lightly Al-doped MgB$_2$ samples. \cite{Pissas02}

\begin{figure}[tbp] \centering
\includegraphics[angle=0,width=8.5cm]{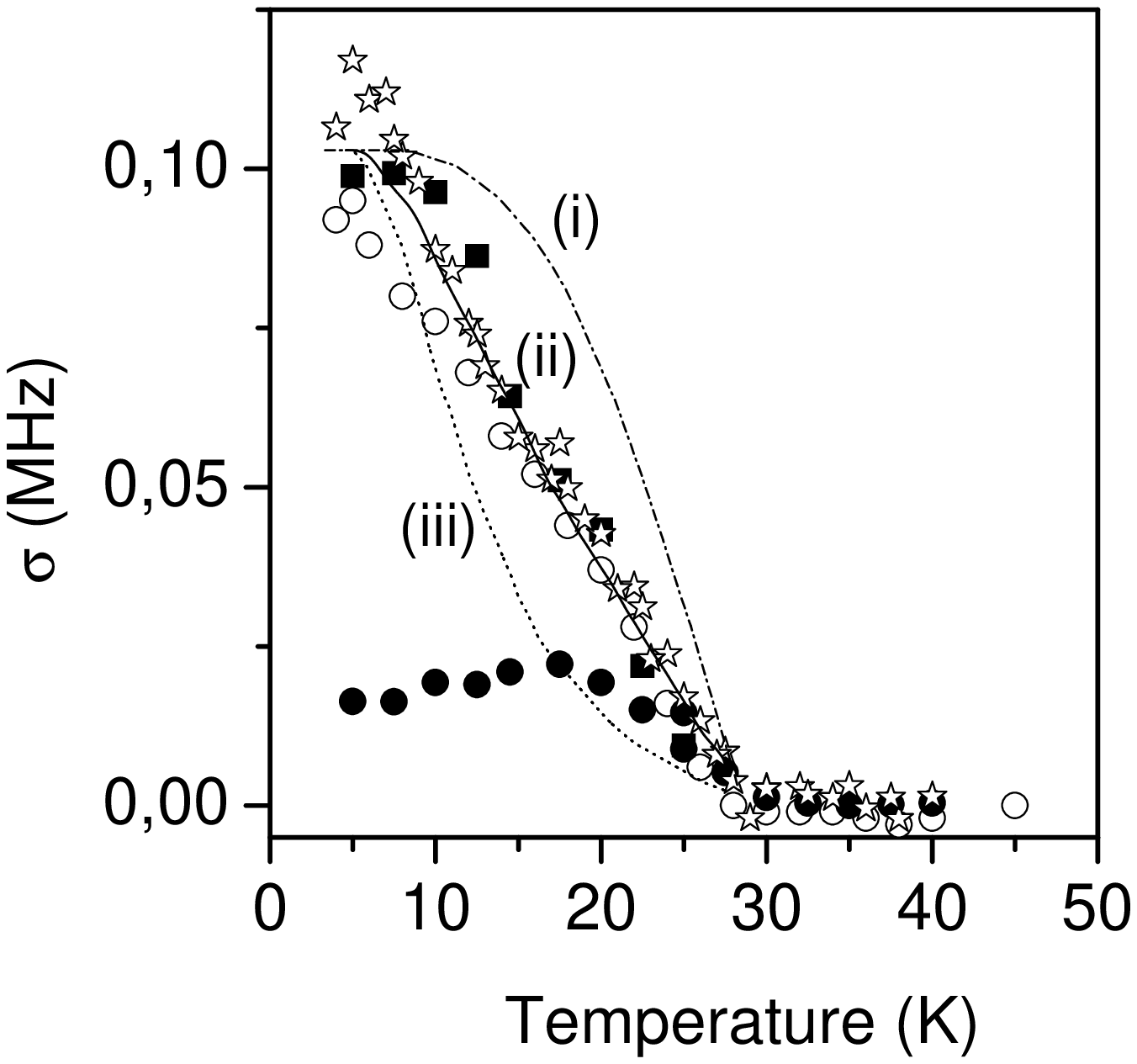}
\caption{
$^{11}$B NMR line width as a function of temperature for pure MgB$_2$ (filled circles are for the normal state and filled squares for the vortex state), MgB$_{1.98}$C$_{0.02}$ (open stars), and MgB$_{1.96}$C$_{0.04}$ (open circles). Lines are theoretical $1/\lambda ^2$ vs. $T$ curves from ref. \cite{Golubov02}, after appropriate scaling for the following cases: (i) conventional BCS superconductivity, (ii) double gaped superconductivity with the magnetic field parallel to the c-axis (in the dirty limit), or to the ab-plane (in the clean limit), and (iii) double gaped superconductivity with the magnetic field parallel to the c-axis (in the clean limit).}
\label{fig4}
\end{figure}

The disappearance of the normal state signal in the carbon doped samples enables the exact calculation of the variance $\sigma $ (second moment) of the frequency distribution in the vortex state as a function of temperature. This is important as the second moment of the NMR frequency distribution, is formally related to the magnetic penetration depth $\lambda $ according to formula $\sigma \approx \gamma_n 0.0608(\phi _0/\lambda ^2)$, where $\gamma _n$ is the nuclear gyromagnetic ratio for boron, and $\phi _0$ is the flux quantum ($=2.07\times 10^{-7}G\cdot cm^2$) \cite{Sonier00}. We note that in powder compounds with strong anisotropy one should take into consideration that $\sigma $ contains contributions from both in-plane $\lambda _{ab}$ and perpendicular $\lambda _c$, which may vary differently with temperature. According to the Ginzburg-Landau theory, clean superconductors with arbitrary gap anisotropy are characterized by a single anisotropy parameter, i.e. $\xi_{ab}/\xi_c=\lambda _{ab}/\lambda _c$. Under this aspect one would expect that for MgB$_2$, $\gamma _\lambda = \gamma _\xi \approx 6$. However, recent theoretical \cite{Kogan02,Miranovic03} and experimental \cite{Cubitt03,Lyard03} studies have shown that due to the two gap character of superconductivity in MgB$_2$,  $\gamma _\lambda $ is nearly isotropic ($\approx 1$) at low temperatures, while it increases to $\approx 2.6$ at $T_c$. It is thus expected that at low temperatures $\sigma $ would reflect a mean $1/\lambda^2$. Here, we would like to stress that there are certain factors which complicate the interpretation of $\sigma $ in terms of the magnetic penetration depth: (i) as pointed out by Brandt \cite{Brandt}, various kinds of random distortions degrade significantly the magnetic field distribution in the vortex state and (ii) in pure MgB$_2$ the high frequency tail of the line shapes overlaps with the normal state signal component, while in the carbon doped samples the tail is strongly supressed. The latter effect might be produced by the establishment of a disorder-driven Bragg glass state, right below $T_{c2}^c$, which partially smooths out the details of the magnetic field distribution. \cite{Pissas03,Ohmichi03} These problems allow only a rough estimation of the penetration depth. For example, the slightly shorter second moment of pure MgB$_2$ at low temperatures in comparison with the carbon doped MgB$_{1.98}$C$_{0.02}$ sample, as shown in Figure \ref{fig4}, may reflect the stronger pinning effect by the carbon dopant. \cite{Kohandel03} Despite the above mentioned shortcomings, we would like to point out some features, which appear to support the two-gap superconductivity scenario for MgB$_2$. First of all, we notice that by assuming a mean $\lambda \approx 1200 \AA$ at low temperatures, we obtain $\sigma \approx 0.11$ MHz, which is not very far from the measured $\sigma-\sigma _N $ values for the three studied samples ($\sigma _N$ is $\sigma$ for the normal state NMR signals), as shown in Figure \ref{fig4}. What is remarkable is that both carbon doped samples exhibit an almost linear temperature dependence of $\sigma $ with a small hump at $\approx 17$ K, which follows relatively well the predicted temperature dependence for $1/\lambda ^2(T)$, as described in a recently proposed two-gap model \cite{Golubov02}. This may be considered as indication that two gap superconductivity is moderately affected at field $23.5$ kOe, in agreement with recent point-contact spectroscopy data which show that at low temperatures the smaller $\pi $ gap survives up to fields $\approx 60$ kOe \cite{Yanson04,Bugoslavsky04}. As can be seen in Figure \ref{fig4}, there is relatively good coincidence between theory and experiment for two cases: (a) with the  magnetic field along the $c$-axis in the dirty limit (predicted $\lambda _{ab}\approx 1057 \AA$), and (b) with the magnetic field along the $ab$-axis in the clean limit (predicted $\lambda _c\approx 392 \AA$). Nevertheless, the latter case can be excluded as it corresponds to order of magnitude broader vortex state NMR signals than those experimentally observed. On the basis of the above $^{11}B$ NMR line shape analysis, it may be concluded that carbon doping increases drastically the coherence length $\xi _{ab}$, while driving the systems from the clean to the dirty limit regime, in agreement with recent observations. \cite{Pissas03,Ohmichi03}

\subsection{$^{11}$B NMR spin-lattice relaxation}

\begin{figure}[tbp] \centering
\includegraphics[angle=0,width=8.5cm]{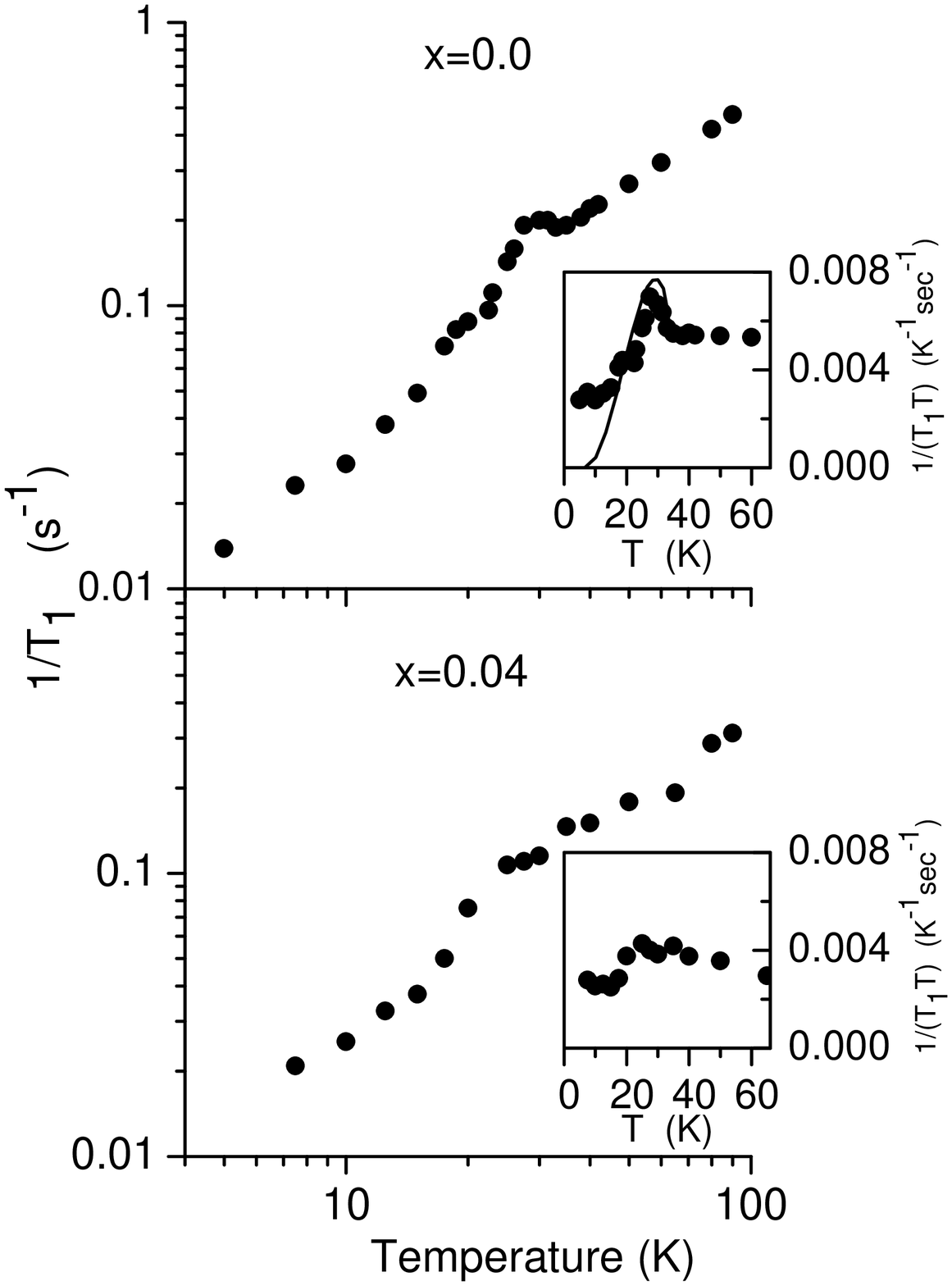}
\caption{
$^{11}$B NMR ($1/T_1$) relaxation rates as a function of temperature for pure MgB$_2$ and carbon doped MgB$_{1.96}$C$_{0.04}$.}
\label{fig5}
\end{figure}

The temperature dependence of $^{11}$B $1/T_1$ for the pure MgB$_2$ and MgB$_{1.96}$C$_{0.04}$  samples is presented in Figure \ref{fig5}. In the case of pure MgB$_2$ for $T>30$K, $T_1$ was measured on the peak of the normal state signal, and for $T\leq 30$K on two different spectral positions: (i) on the peak (saddle point) of the signal component from the vortex state, and (ii) on the peak of the signal component from the normal state, by applying the two exponential analysis described by Kotegawa et al. \cite{Kotegawa01,Kotegawa02}

\begin{figure}[tbp] \centering
\includegraphics[angle=0,width=8.5cm]{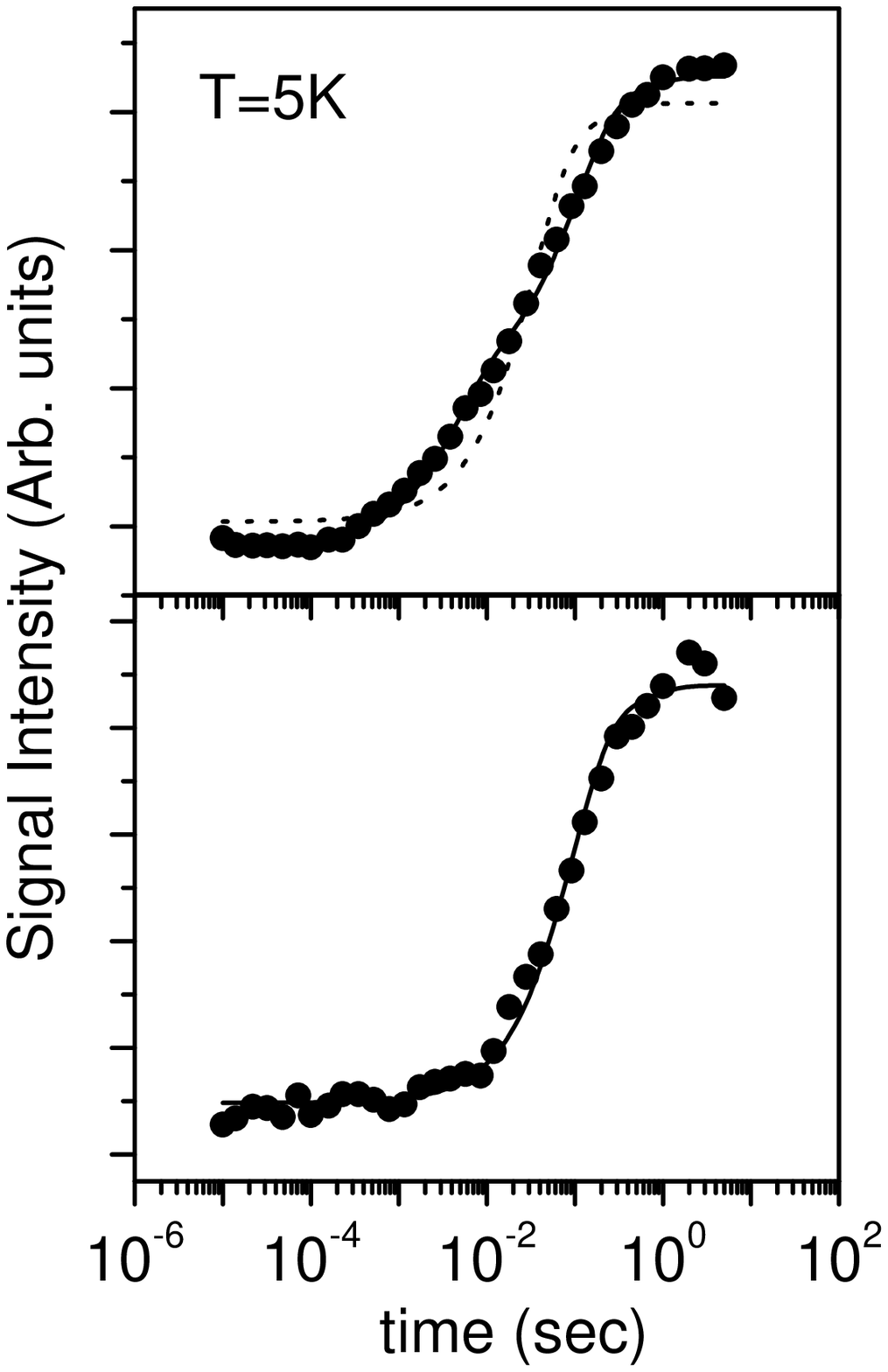}
\caption{
$^{11}$B NMR saturation recovery curves of pure MgB$_2$ at $T=5K$, for measuring $T_1$ of the normal signal component (uper panel) and of the vortex signal component (lower panel). The vortex state signal was nicely fitted with a single exponential analysis. Fit of the normal signal component was only possible by considering a double $T_{1,n}$ and $T_{1,v}$ analysis, where  $T_{1,v}$ was set equal to the $T_1$ value of the vortex state signal component (the dot-line is the best single exponential fit).}
\label{fig6}
\end{figure}

The biexponential nature of the nuclear magnetization recovery curves in this latter case is clearly demonstrated in Figure \ref{fig6}, which exhibits $^{11}$B NMR saturation recovery curves at $T=5$K for the normal state signal component (upper panel) and the vortex state signal component (lower panel). While the saturation recovery curve from the vortex state signal can be fitted by considering a single $T_{1,v}=3.5$ s, in the case of the normal state signal component, a nice fit is only obtained by considering two relaxation times: a long one $T_{1,v}=3.5$ s, which follows the temperature dependence of $T_1$ in the vortex state, and a short one $T_{1,n}= 0.2$ s. Similar results have been reported in refs. \cite{Kotegawa02,Baek02} Apparently, the short $T_{1,n}$ component belongs to the normal state NMR signal component from grains in the normal state, and will be neglected in the following. However, the corresponding $1/(T_{1,n}T)$ values below $T_c$ are enhanced over the Korringa value of the normal phase, as also pointed out in ref. \cite{Baek02}, a phenomenon which is not yet understood. 

The evolution of the relaxation rate $1/T_1$ as a function of temperature in the mixed superconducting state is shown in the upper panel of Figure \ref{fig5}. It is observed that above $T_c\approx 30$K, $(1/(T_1T))$ is constant, with a value $\approx 0.0056$ $K^{-1}s^{-1}$, following a Korringa temperature dependence. Right below $T_c$, $(1/(T_1T))$ forms a peak centered at $T\approx 28$K. This peak is obviously attributed to the coherence (Hebel-Slichter) peak \cite{HebelSlichter59} and indicates a fully gaped s-wave pairing state. This effect has been recently demonstrated by Kotegawa et al. in a $^{11}B$ NMR study of lightly doped Mg$_{1-x}$Al$_x$B$_2$ samples. \cite{Kotegawa02} The solid curve in the inset of Figure \ref{fig5} is the T dependence of $1/(T_1T)$, by using the typical $s$-wave model, where a phenomenological energy broadening function is assumed to be of a rectangle type with a width $2\delta $ and a height $1/2\delta $ as presented by Hebel. \cite{Hebel59} By considering a superconducting gap size of $2\Delta /k_BT_c\approx 4.5$ and $\delta /\Delta(0)\approx 1/3$ in agreement with Kotegawa at al. \cite{Kotegawa01}, a nice fit is obtained down to $15$ K; however, $(1/(T_1T))$ appears to saturate at lower temperatures. The origin of this effect is not quite clear. However, the opening of the second gap as possible origin of the deviation from the typical BCS behaviour is excluded as it does not gives the almost horizontal low temperature slope of the $(1/(T_1T))$ vs $T$ curve. Deviation from the BCS behaviour at low temperatures has been already reported in both conventional isotropic \cite{Fite66,Silbernagel66} and anisotropic type II superconductors. \cite{Bulaevskii93,Curro00} This phenomenon has been attributed to contribution in the relaxation mechanism from the vortex dynamics \cite{Bulaevskii93,Curro00}, which becomes predominant at low temperatures, where the BCS contribution to the relaxation rates is very small. On the basis of this argumentation $(1/(T_1))$ in the mixed superconducting state of MgB$_2$ is expected to be given by $1/T_1=1/T_{1,v}+1/T_{1,vm}$, where $1/T_{1,v}$ and $1/T_{1,vm}$ is the contribution from quasiparticle scattering and the vortex motion respectively. Considering the vortex fluctuations, we have $1/T_{1,vm}=\gamma ^2h_\perp ^2\cdot \tau _c/(1+\omega _L^2\tau _c^2)$, where $\omega _L$ is the Larmor frequency and $h_\perp $ is the fluctuating field in the perpendicular direction.
In the case of ($\omega _L\tau _c)\gg 1$ and by considering that $h_\perp \sim T$, the above formula becomes $1/T_{1,vm}\approx \gamma ^2h_\perp /\omega _L^2\tau _c\sim T/\tau _c$, and $1/T_1$ shows at low temperatures the linear $T$-dependence as in Figure \ref{fig5}.

At the same time, even a small ammount of carbon doping like $x=0.04$ leads to significant suppression of both the coherence peak and of the rapid fall of $1/(T_1T)$ below $T_c$, as observed in the lower pannel of Figure \ref{fig5}. First of all we would like to notice that the opening of the superconducting gap suppresses $1/T_1$, while the sudden increase of $N(E_F)$ at $T_c$ acts in the reverse way ($1/(T_1T)\propto N^2(E_F)$), so the coherence peak in $1/(T_1T)$ may be considered as the result of the competition between these two factors. Hence, any change in the gap strength or $N(E_F)$, would affect directly the coherence peak and the slope of the $1/(T_1T)$ vs $T$ curve below $T_c$. It is also notable that (i) according to the literature non-magnetic inpurity- like carbon -  scattering smears out the gap anisotropy and enhances the coherence peak \cite{Masuda62,Williamson73,Choi03}, and (ii) the two-gap behaviour is retained upon light carbon doping. \cite{Papagelis03a,Schmidt03,Samuely03,Ohmichi03} It is thus straightforward to conclude that the supression of the coherence peak is not caused by impurity scattering, but is rather consequence of the shrinking of the $\sigma $ hole bands with carbon, i.e. electron doping.

\section{Conclusion}

In conclusion, we have shown that even a small substitution of boron with carbon in MgB$_2$ leads to substantial increase of $H_{c2}^c$ and unveils the magnetic field distribution in the vortex state, as reflected in the $^{11}B$ NMR frequency distribution. The temperature dependence of the second moment of the NMR signal $\sigma $ vs. $T$ shows a clear deviation from the BCS theory of the penetration depth $\lambda$ and correlates nicely with a two gap model for $\lambda $. The $H_{c2}^c$ enhancement indicates drastic reduction of $\xi _{ab}$ and passage from the clean to the dirty limit with light carbon doping. According to the spin-lattice relaxation measurements, in pure MgB$_2$ a strong coherence peak is observed close to $T_c$, followed by rapid decrease in the $1/(T_1T)$ vs. $T$ plot, which is indicative of the s-type nature of superconductivity. In carbon-doped systems, both effects faint out, presumably because of gradual electron filling and shrinking of the $\sigma $ hole-type cylindrical Fermi surfaces.

\end{document}